\documentclass[11pt]{article}

\usepackage[final]{acl}

\usepackage{times}
\usepackage{latexsym}

\usepackage[T1]{fontenc}

\usepackage[utf8]{inputenc}

\usepackage{microtype}

\usepackage{inconsolata}

\usepackage{graphicx}

\title{Towards Human-Centered RegTech: Unpacking Professionals' Strategies and Needs for Using LLMs Safely}

\author{
 \textbf{Siying Hu\textsuperscript{1}},
 \textbf{Yaxing Yao\textsuperscript{2}},
 \textbf{Zhicong Lu\textsuperscript{3}},
\\
\\
 \textsuperscript{1}City University of Hong Kong,
 \textsuperscript{2}Johns Hopkins University,
 \textsuperscript{3}George Mason University
 }

\begin{document}
\maketitle
\begin{abstract}
Large Language Models are profoundly changing work patterns in high-risk professional domains, yet their application also introduces severe and underexplored compliance risks. To investigate this issue, we conducted semi-structured interviews with 24 highly-skilled knowledge workers from industries such as law, healthcare, and finance. The study found that these experts are commonly concerned about sensitive information leakage, intellectual property infringement, and uncertainty regarding the quality of model outputs. In response, they spontaneously adopt various mitigation strategies, such as actively distorting input data and limiting the details in their prompts. However, the effectiveness of these spontaneous efforts is limited due to a lack of specific compliance guidance and training for Large Language Models. Our research reveals a significant gap between current NLP tools and the actual compliance needs of experts. This paper positions these valuable empirical findings as foundational work for building the next generation of Human-Centered, Compliance-Driven Natural Language Processing for Regulatory Technology (RegTech), providing a critical human-centered perspective and design requirements for engineering NLP systems that can proactively support expert compliance workflows.
\end{abstract}

\section{Introduction}
The rapid development of Large Language Models (LLMs) is sparking a productivity revolution in knowledge-intensive industries \cite{eloundou2023gpts,weber2024significant,li2023sheetcopilot}. Professionals across various fields, from drafting legal documents to analyzing financial data, are increasingly using these tools to enhance their efficiency and the quality of their work outcomes \cite{cheong2024not,kim2024financial}. However, in these highly regulated industries that frequently handle sensitive information, the widespread adoption of LLMs also brings unprecedented compliance challenges \cite{rathod2025privacy}. These risks stem from multiple sources, including potential data privacy breaches \cite{yan2024protecting}, intellectual property infringement \cite{yu2023codeipprompt}, and issues with the accuracy of model-generated content \cite{boutadjine2025human} (i.e., hallucinations), which could lead to severe legal, ethical, and reputational repercussions. For instance, there have been reports of corporate employees inadvertently inputting confidential information such as company source code and internal meeting records into public LLM tools \cite{dora2025hidden}.

Although existing research has addressed general privacy and security concerns of LLMs \cite{das2025security,yan2024protecting}, little is known about how highly-skilled knowledge workers specifically perceive, assess, and respond to compliance risks within their professional workflows. These experts possess deep domain knowledge and extensive practical experience; their compliance judgments are not merely about following rules but involve professional decision-making in complex and ambiguous situations. The ``ad-hoc strategies'' they spontaneously form to mitigate risks represent a treasure trove of insights waiting to be explored, crucial for understanding the human-computer interaction bottlenecks in real-world applications.

The core argument of this study is that to build truly effective and trustworthy NLP-based Regulatory Technology (RegTech) systems \cite{cejas2023nlp,zhang2025ai}, we need first proceed from a Human-Centered perspective \cite{karpen2021designing,saltarella2024translating} to deeply understand the real-world dilemmas and practical wisdom of end-users. Many current technical solutions overlook this critical human dimension, attempting to solve socio-technical problems with purely technical means, which results in a disconnect between technological capabilities and actual user needs \cite{ackerman2000intellectual}. Through an interview study with experts, this research aims to bridge this gap and provide bottom-up, empirical evidence from the front lines to inform system design.

To this end, we propose and aim to answer the following three research questions:
(1) What compliance risks do knowledge workers perceive when using LLMs to complete textual work?
(2) How do knowledge workers mitigate these compliance risks?
(3) What challenges and requirements do knowledge workers have when mitigating these risks?

By answering these questions, this short paper contributes to the HCI and NLP communities with a detailed empirical account of how highly-skilled workers navigate the compliance risks of LLMs. More importantly, we distill these findings into a series of implications for future system design, aiming to provide a solid empirical foundation and a design blueprint for developing the next generation of trustworthy NLP tools that can proactively support compliance.

\section{Related Work}

This research is positioned at the intersection of three academic domains: the application of Large Language Models (LLMs) in knowledge work, the technical risks and governance of LLMs, and Human-Computer Interaction studies on technology adoption and workarounds. Our contribution lies in connecting these domains to focus on a critical, overlooked link: the subjective perception and practical mitigation of compliance risks by highly-skilled experts in real-world work contexts.

LLMs in Knowledge Work. A substantial body of research has confirmed that LLMs can significantly enhance the productivity of knowledge workers \cite{li2023sheetcopilot,weber2024significant}. The literature documents use cases across various fields such as programming, legal writing, academic research, and content creation \cite{meyer2023chatgpt,siino2025exploring,zhang2024experimenting,kasneci2023chatgpt} , demonstrating their immense potential in accelerating information retrieval, drafting initial texts, and sparking creativity \cite{chakrabarty2024art}. However, these studies predominantly focus on efficiency and output quality, seldom delving into the compliance challenges that coexist with these productivity gains in high-stakes domains.

Technical Risks and Governance of LLMs. Another active research area analyzes LLM risks from a technical and policy perspective. Researchers have identified numerous technical risks, including data privacy leakage, model bias, content hallucination, and intellectual property infringement \cite{yan2024protecting,yu2023codeipprompt}. Correspondingly, they have proposed technical mitigation solutions like differential privacy, data de-identification, model watermarking, and content provenance \cite{lukas2024analyzing} , as well as governance frameworks such as AI ethics guidelines and industry regulations \cite{wirtz2022governance}. These top-down solutions are vital for building a macro-level safe AI ecosystem. However, they are often disconnected from the actual workflows of frontline users and fail to address the here and now compliance dilemmas experts face in specific tasks, thus leaving the socio-technical gap that we address.

Technology Adoption and Workarounds in HCI. The HCI field has long been interested in how users adopt, adapt, and even creatively misuse new technologies \cite{de2006misuse,gugenheimer2020exploring}. Concepts like workaround and appropriation are used to describe the spontaneous adjustments users make to fit technology into their specific work contexts \cite{majchrzak2000technology}. This research reveals the tension between a technology’s designed affordances and professional's workplace needs \cite{duan2024technology}. However, in the context of LLMs, there is a research gap concerning the systematic workarounds that highly-skilled experts perform specifically for compliance-driven motives. Our work aims to fill this gap by exploring experts’ ad-hoc strategies to provide practice-grounded insights for designing truly human-centered compliance technology.

\section{Method}
To investigate the compliance risk perceptions and practices of knowledge workers using Large Language Models, we conducted a qualitative study \cite{malterud2016sample}. We recruited 24 highly-skilled knowledge workers, with a balanced gender distribution of 12 men and 12 women. These participants came from eight knowledge-intensive industries, including academia, law, healthcare, technology, and commerce. Recruitment occurred from July to September 2024, and we used word-of-mouth and a screening process to ensure participants had at least three years of relevant work experience. We employed an extreme case sampling strategy \cite{onwuegbuzie2007sampling}, focusing on high-compliance-pressure domains like law and healthcare, as the risks and challenges in these fields are more representative of the core issues.

We conducted semi-structured interviews in Chinese to ensure precise and nuanced communication with the non-native English-speaking participants. Each interview lasted between 45 and 60 minutes. Before the interviews, we provided a short primer to establish a consistent baseline understanding of Large Language Models across all participants. The core of the interview was to elicit participants' recollections of recent tasks completed with model assistance and to probe deeply into their considerations, actions, and challenges regarding privacy, security, and compliance.

With consent, all interviews were audio-recorded and transcribed. We used thematic analysis to analyze the data \cite{castleberry2018thematic}. The analysis involved a two-step process. In the first step, two researchers used open coding to inductively identify concepts from the transcripts related to risk perceptions, mitigation practices, and organizational support. In the second step, we synthesized these initial codes into higher-level themes. For instance, codes like “delete real name” were grouped into the strategic theme of ``Intentional Misleading''. The entire analysis process involved regular discussions and cross-validation among researchers to ensure the reliability of the findings.

\section{Findings}
Our analysis revealed a significant disconnect between experts operational behavior of large language models (LLMs) and the compliance prerequisites of professional environments. Participants' feedback points to critical deficiencies in model governance, explainability, and the broader AI supply chain, forcing them into high-effort, manual mitigation patterns.

Participants' perceived compliance risks are primarily rooted in the models' architectural limitations concerning data governance and lifecycle management. The foremost risk, information security and privacy leakage, stems from the potential for model memorization and training data contamination. Participants expressed valid concerns that sensitive inputs could be integrated into subsequent model versions or exposed through inversion attacks, creating direct liabilities under data protection regulations like GDPR and CCPA. As one management consultant (P12) stated, this transforms the tool into a source of unacceptable risk: \textit{``It's like throwing my trade secrets into a black box... If something goes wrong, I lose not just the client, but the credibility of my entire career.''} A second critical risk is intellectual property infringement, driven by the models' lack of output provenance and data attribution. This creates a dual threat: outputs may constitute derivative works of copyrighted material, and users' novel prompts could be absorbed as training data without attribution. Finally, the risk of output inaccuracy (i.e., "hallucinations") was framed as a failure of factual grounding, posing a direct threat of professional liability.

In the absence of built-in technical safeguards, participants are forced to implement manual, user-side control frameworks. One common strategy is proactive input sanitization, a rudimentary form of data minimization and masking. This involves manually redacting sensitive entities and reducing prompt specificity to limit the model's contextual inference capabilities. The other primary strategy is an intensified Human-in-the-Loop verification process. By adopting a ``zero-trust'' policy, participants treat all outputs as unverified drafts requiring a full post-processing validation layer. This manual audit function, as one analyst (P5) highlighted, fundamentally undermines the model's efficiency proposition: \textit{``This isn't improving efficiency; it's just adding to my 'auditing' workload.''}

These user-driven efforts are ultimately insufficient due to systemic challenges inherent in the current AI ecosystem. The core issue is the lack of model explainability and causal traceability. Current LLMs do not provide the necessary mechanisms for data-to-output attribution, making it impossible to conduct a formal risk assessment or audit. This technical opacity creates a normative vacuum in regulation, as existing legal frameworks for liability cannot be effectively applied. As one corporate counsel (P4) noted, this forces them to operate in a state of profound uncertainty: \textit{``The company has not yet formed a formal AI use policy, so we are operating in a 'gray area'.''} This culminates in a fractured chain of responsibility within the AI supply chain (developer → provider → deployer → user). Without a clear delineation of liability, and with the technical substrate for auditability being absent, the end-user is left feeling disproportionately exposed and powerless

\section{Implications}
Our findings not only reveal user-level dilemmas but also directly provide critical design implications for the interdisciplinary research area of Human-Centered NLP and HCI. We summarize these implications into three aspects.

Transparency by Design, from Black Box to Proactive Guidance. Our research revealed that experts feel confused due to a lack of targeted training. Therefore, future NLP systems should not be merely passive tools but should act as proactive guides. Systems should be able to provide real-time, context-aware compliance risk alerts tailored to the user's industry and specific task. For example, when a legal professional inputs case information, the system could automatically highlight potential client privacy risks and link to relevant ethical codes.

Interaction Design that Supports Professional Judgment and Values. The experts demonstrated a strong sense of professional ethics and responsibility, yet existing tools fail to support their value judgments. This inspires us to adopt the principles of Value-Sensitive Design, embedding values such as fairness, privacy, and accountability into the system's interaction. For instance, systems should provide clear feedback channels and grant users final review and modification rights over AI suggestions, thereby supporting rather than replacing expert judgment.

Practical NLP Tools and Scaffolding to Empower Users. To empower users, future systems need to provide concrete automation tools to reduce the compliance burden. This includes developing NLP modules that can automatically identify and de-sensitize personally identifiable information (PII) or trade secrets as the user inputs them. Furthermore, systems can offer a scaffolding-style interactive workflow, guiding users through high-risk tasks with compliance checklists and standard operating procedures to ensure consistency and adherence to regulations.

\section{Conclusion}
Through interviews with 24 highly-skilled knowledge workers, this study revealed the compliance risks they perceive when integrating Large Language Models, including data leakage, intellectual property issues, and output uncertainty. We found that they respond with ad-hoc strategies such as distorting inputs, but the effectiveness of these efforts is limited by a lack of systemic support, regulatory lag, and unclear responsibilities. Our core contribution is an empirically grounded report on the needs of frontline experts, which we argue should serve as the starting point for building the next generation of human-centered, compliance-driven NLP tools. Our findings offer concrete and actionable guidance for designing NLP systems that are trustworthy, compliance-aware, and capable of proactively supporting experts. Future work should expand the scope of research to more industries and incorporate quantitative methods for a more comprehensive risk assessment. Furthermore, exploring the new compliance challenges that arise when AI participates in collaboration as more agentic entities (Agents) will be an important research direction.

\section{Limitations and Future Work}
As a exploratory study, our work has limitations that suggest future research directions. First, our qualitative sample of 24 experts, while providing rich data, is not statistically generalizable; future large-scale quantitative surveys are needed to validate our findings across diverse contexts. Second, our reliance on self-reported data may be subject to recall and social desirability biases. This could be complemented by observational methods like log analysis to capture actual user behavior. Finally, our individual-level focus simplifies the complex organizational context. Future research should adopt a more ecological perspective to investigate the distribution of responsibilities among users, organizations, and technology providers, especially exploring the new compliance challenges that will arise as AI emerges in more agentic forms.

\bibliography{Main}

\appendix



\end{document}